# UTILIZAREA INDICATORILOR FINANCIARI PENTRU A IDENTIFICA FIRMELE ROMANESTI CU DIFICULTATI

# USING FINANCIAL RATIOS TO IDENTIFY ROMANIAN DISTRESSED COMPANIES

Madalina Ecaterina ANDREICA, PhD Student, ASE, madalina.andreica@gmail.com
Mugurel Ionut ANDREICA, PhD Student, U.P.B., mugurelionut@gmail.com
Marin ANDREICA, PhD, ASE Bucharest, marinandreica@yahoo.com

**Rezumat**
    In contextual actualei crize financiare, cand din ce in ce mai multe firme se confrunta cu situatii de faliment sau de insolvabilitate, lucrarea isi propune sa gaseasca metode pentru a identifica pe baza indicatorilor financiari, firmele aflate in dificultate. Studiul se va axa pe identificarea unui grup de de companii romanesti listate la bursa, pentru care au fost disponibile date financiare pentru anul 2008. Pentru fiecare societate s-a calculat un set de 14 indicatori financiari, ce a fost apoi utilizat intr-o analiza a componentelor principale, urmata de o analiza cluster, intr-un model logit, precum si intr-un model de arbori de clasificare de tip CHAID.
**Cuvinte cheie**: firme cu dificultati, indicatori financiari, cluster, CHAID, modelul logit

**Abstract**
    In the context of the current financial crisis, when more companies are facing bankruptcy or insolvency, the paper aims to find methods to identify distressed firms by using financial ratios. The study will focus on identifying a group of Romanian listed companies, for which financial data for the year 2008 were available. For each company a set of 14 financial indicators was calculated and then used in a principal component analysis, followed by a cluster analysis, a logit model, and a CHAID classification tree.

## 1. INTRODUCTION

The financial crisis has already thrown many financially strong companies out of business all over the world. All this happened because they were not able to face the challenges and the unexpected changes in the economy. In Romania, for example, a study made by Coface Romania and based on the data provided by the National Trade Register Office, stated that around 14.483 companies were forced into bankruptcy by the end of the year 2008 or became financially distressed when they were not able to pay their financial obligations due to inadequate cash flows.

Looking at the above situation, we realise how important it is to understand the reasons behind the collapse of a company. Knowing these reasons might hinder a company from being financially distress and early actions could be taken as a precaution.

The study of this paper will focus on identifying a group of distressed and non-distressed Romanian listed companies in 2008, for which data were available, and then to predict financial distressed companies by using **the Logistic model**. For each company a set of 14 financial ratios were calculated and then used in the process of identifying and predicting the distressed companies. The study also includes **a principal component analysis**, in order to better estimate the importance of each financial ratio included in the study, as well as several methods of classification such as a **cluster analysis** and **the CHAID classification tree method.**

## 2. LITERATURE REVIEW

The first step in the evolution of the quantitative firm failure prediction model was taken by **Beaver** (1966), who developed a dichotomous classification test based on a simple *t*-test in a univariate framework. He used individual financial ratios from 79 failed and non-failed companies that were matched by industry and assets size in 1954 to 1964 and identified a single financial ratio – *Cash flow/ Total Debt* as the best predictor of corporate bankruptcy.

Beaver's study was then followed by **Altman** (1968), who suggested a Multivariate Discriminant Analysis (**MDA**). By utilizing 33 bankrupt companies and 33 non-bankrupt companies over the period 1946 – 1964, five variables were selected most relevant in predicting bankruptcy. These were *Working Capital to Total Assets*, *Retained Earnings to Total Assets, Earnings before Interest and Taxes to Total Assets, Market Value of Equity to Book Value of Total Debt* and *Sales to Total Assets*. The MDA model was able to provide a high predictive accuracy of 95% one year prior to failure. For this reason, MDA model had been used extensively by researchers in bankruptcy research (Altman, Haldeman and Narayanan, 1977; Apetiti, 1984; Izan, 1984, Micha, 1984; Shirata, 1998; Ganesalingam and Kumar, 2001).

However, Eisenbeis (1977), Ohlson (1980) and Jones (1987) found that there were some inadequacies in MDA with respect to the assumptions of normality and group dispersion. The assumptions were often violated in MDA and this may biased the test of significance and estimated error rates.

**Logit analysis** which did not have the same assumptions as MDA was made popular by **Ohlson** (1980). He used 105 bankrupt companies and 2058 non-bankrupt companies from 1970 to 1976. The results showed that *size,* financial structure (*Total Liabilities to Total Assets*), performance and current liquidity were important determinants of bankruptcy. In the logit analysis, average data is normally used and it is considered as a single period model. Hence, for each non-distressed and distressed company, there is only one company-year observation. The dependent variable is categorized into one of two categories that is distressed or non-distressed.

In 2004, two econometric problems with the single period logit model were discussed. (Hillegeist, 2004). First, is the sample selection bias that arises from using only one, non-randomly selected observation for each bankrupt company, and second, the model fails to include time varying changes to reflect the underlying risk of bankruptcy. This will induce crosssectional dependence in the data. Shumway (2001) demonstrated that these problems could result in biased, inefficient, and inconsistent coefficient estimates. To overcome these econometric problems, **Shumway** (2001) predicted bankruptcy using **the hazard model** and found that it was superior to the logit and the MDA models. This particular model is actually a multi-period logit model because the likelihood functions of the two models are identical. For this reason, the discrete-time hazard model with time-varying covariates can be estimated by using the existing computer packages for the analysis of binary dependent variables. The main particularities of the hazard model consist in the facts that firm specific covariates must be allowed to vary with time for the estimator to be more efficient and a baseline hazard function is also required, but which can be estimated directly with macroeconomic variables to reflect the radical changes in the environment.

In recent years many types of heuristic algorithms such as **neural networks** and **decision trees** have also been applied to the bankruptcy prediction problem and several improvements in the financial distress prediction were noticed. Research studies on ANN for bankruptcy prediction started in 1990 like Bell et al.(1990), Tam and Kiang(1992),Wilson et al,(1992), Coats and Fant (1993),Udo(1993), Fletcher and Goss (1993), Altman et al.(1994), Boritz and Kennedy (1995), Back et al.(1996), Etheridge and Sriram (1997), Yang et al.(1999), Fan and Palaniswami(2000), Atiya(2001) used to forecast financial distress for bank and other business, and are still active now. Neural networks are non-linear architectures, so that they are able to discriminate patterns which are not linearly separable and do not require data to follow any specific probability distribution. Neural Networks have been discovered to be better classifiers than discriminant analysis methods in a number of works based on financial data from American firms (Odom and Sharda 1990; Tam and Kiang 1990,1992; Coats and Fant, 1993; Wilson and Sharda 1994).

The main disadvantage of neural network models, however, consists in the difficulty of building up a neural network model, the required time to accomplish iterative process and the difficulty of model interpretation. Compared to neural networks, decision tree is not only a non-linear architecture, which is able to discriminate patterns that are not linearly separable and allow data to follow any specific probability distribution, but also plain to interpret its results, require little preparation of the initial data and perform well with large data in a short time.

**Zheng and Yanhui** (2007) used **decision tree** methodologies for corporate financial distress prediction in their study. The authors presented the advantages of using CHAID decision trees in comparison to a neural network model, which is complicated to build up and to interpret or to a statistic model such as multivariate discriminate regression and logistic regression, where the patterns need to be linearly separable and samples are assumed to follow a multivariate normal distribution. Their study focused on 48 failed and continuing Chinese listed companies in the period 2003 – 2005. The following variables embodied most information for predicting financial distress: *Net Cash Flow from Operating Activity as a percentage of Current Liabilities, Return Rate on Total Assets, Growth rate of Total Assets* and *Rate on Accounts Receivable Turnover*. They also noticed that it is not appropriate to use financial information to predict financial distress ahead of four years. However, the results supported by the test study showed that decision trees was a valid model to predict listed firms' financial distress in China, with a 80% probability of correct prediction.

## 3. RESEARCH DESIGN

### 3.1. Data description

For this study, financial information for the year 2008 was collected for a total sample of 55 Romanian listed companies, from the Bucharest Stock Exchange. The companies were divided into two categories, namely "healthy" and "unhealthy" companies, also called non-distressed and distressed companies. Since there is no standard definition for a "distressed" company, we considered a company to be "unhealthy" in case it had losses for at least two consecutive years or in case it had unpaid taxes or any other debts for at least two consecutive years. In the sample there were 8 companies with losses for at least two years and 4 companies with unpaid taxes for at least two years. However, there were also 6 companies with losses for the year 2008, which were also included in the "unhealthy" category, since this can also be considered a weaker but still possible sign of distress in an uncertainty situation. To summarize, from the total sample of 55 companies, 18 were classified as "distressed" and 37 as "non-distressed".

### 3.2. Financial ratios

The selection of the main set of financial ratios was mainly based on the previous results presented in the related work, but also restricted to the financial data provided by the Bucharest Stock Exchange. There were 14 financial ratios used in this study, which are presented in the table below. They were grouped into 5 distinct categories, based on issues such as profitability, solvency, asset utilization, growth ability and company size.

| Category | Code | Financial ratios | Definition |
|---|---|---|---|
| Profitability | I1 | Profit Margin | Net Profit or Loss / Turnover *100 |
| | I2 | Return on Assets | Net Profit or Loss / Total Assets *100 |
| | I3 | Return on Equity | Net Profit or Loss / Equity *100 |
| | I4 | Profit per employee | Net Profit or Loss / number of employees |
| | I5 | Operating Revenue per employee | Operating revenue / number of employees |
| Solvency | I6 | Current ratio | Current assets / Current liabilities |
| | I7 | Debts on Equity | Total Debts / Equity *100 |
| | I8 | Debts on Total Assets | Total Debts / Total Assets *100 |
| Asset utilization | I9 | Working capital per employee | Working capital / number of employees |
| | I10 | Total Assets per employee | Total Assets / number employees |
| Growth ability | I11 | Growth rate on net profit | (Net P/$L_1$ - Net P/$L_0$) / Net P/$L_0$ |
| | I12 | Growth rate on total assets | (Total Assets$_1$ – Total Assets$_0$) / Total Assets$_0$ |
| | I13 | Turnover growth | (Turnover$_1$- Turnover$_0$) / Turnover$_0$ |
| Size | I14 | Company size | ln (Total Assets) |

### 3.3. Models and methodologies

#### 3.3.1. Principal component analysis

**Principal component analysis (PCA)** is a way of identifying patterns in data, and expressing the data in such a way as to highlight their similarities and differences. Since patterns in data can be hard to find in data of high dimension, where graphical representation is not available, PCA is a powerful tool for analysing data. The other main advantage of PCA is that once you have found these patterns in the data, and you compress the data by reducing the number of dimensions, without much loss of information. By dimensionality reduction in a data set only those characteristics of the data set that contribute most to its variance are kept. PCA offers a convenient way to control the trade-off between loosing information and simplifying the problem by reducing the dimension of the representation.

### 3.3.2. Cluster analysis

**Cluster analysis** or **clustering** is the assignment of a set of observations into subsets (called *clusters*) so that observations in the same cluster are similar in some sense. Clustering is a method of unsupervised learning and a common technique for statistical data analysis used in many fields. Hierarchical cluster analysis contains agglomerative methods and divisive methods that finds clusters of observations within a data set. The divisive methods start with all of the observations in one cluster and then proceeds to split (partition) them into smaller clusters. The agglomerative methods begin with each observation being considered as separate clusters and then proceeds to combine them until all observations belong to one cluster.

In practice, the agglomerative methods are of wider use. On each step, the pair of clusters with smallest cluster-to-cluster distance is fused into a single cluster. The most common algorithms for hierarchical clustering are: *the nearest neighbour* (or the single linkage clustering, where the distance between two clusters is computed as the minimal object-to-object distance), the *farthest neighbor method* (or the complete linkage clustering, where the distance between two clusters is computed as the maximal object-to-object distance), *the average linkage clustering*, where the distance between two clusters is computed as the average distance between objects from the first cluster and objects from the second cluster, *the average group linkage*, where the distance between two clusters is computed as the distance between the average values also known as centroids and *the Ward's linkage*, where the distance between two clusters is computed as the increase in the "error sum of squares" (ESS) after fusing two clusters into a single cluster. The outcome is represented graphically as a dendogram.

### 3.3.3. CHAID

Chi-square automatic interaction detection (CHAID) was originally designed to handle categorical attributes only. For each input attribute, CHAID finds the pair of values that is least significantly different with respect to the target attribute. The significant different is measured by the *p*-value obtained from a statistical test. The statistical test used depends on the type of the target attribute. If the target attribute is continuous, an F-test is used, if it is categorical, then a Pearson chi-square test is used, if it is ordered, then a likelihood-ratio test is used. For each selected pair, CHAID checks if *p*-value obtained is greater than a certain merge threshold. If the answer is positive, it merges the values and searches for an additional potential. The advantage of a CHAID classification tree is that it generates classification rules for the analyzed sample.

### 3.3.4. The Logistic Model

The **logistic model** is a conditional probability model that uses maximum likelihood estimation to provide the conditional probability of a firm belonging to a certain group given the values of the independent variables for that firm. It is a single-period classification model (Shumway, 2001) decribed by the function:

$$P(y_i = 1) = \frac{1}{1 + e^{-x_i \beta}}$$

An important issue in using binary state prediction models such as logit analysis is the selection of the cutoff probability which determines the classification accuracy. In order to classify an observation into one of the two groups, the estimated probability from the logit model is compared to a pre-determined cutoff probability. If the estimated probability is below the cutoff, the observation is classified as an inferior performer and if the estimated probability is above the cutoff, it is placed in the superior performer group.

## 4. ANALYSIS OF RESULTS

The analysis was made for the sample of 55 Romanian listed companies, by using only the financial data of the year 2008. The analysis consisted in applying a principal component analysis followed by a hierarchical cluster analysis, a CHAID decision tree model and a logistic model in order to classify the "healthy" and "unhealthy" Romanian listed companies as well as to identify the most significant financial ratio that contribute most to financial distress prediction.

**PRINCIPAL COMPONENT ANALYSIS:**

We used SPSS 13.0 software for the set of data containing 14 financial ratios for all 55 Romanian listed companies, out of which 37 were "healthy" and 18 "unhealthy". The correlation matrix indicated some strong correlations between the next financial ratios: **I1 and I2** (86%), **I5 and I10** (81.8%), **I2 and I4** (81.7%), **I1 and I4** (77.5%) and **I1 and I6** (75%). In order to reduce the dimension of the initial set of data and also to identify which variables should be kept in order to loose as little information as possible, we applied the principal component analysis. After several tests and after excluding one by one several financial ratios that were most correlated between each other, we reached just two principal components with a total gain of information of **75%**. The results are presented below and indicate that the first principal component is best explained by I1, I2, I4, I6 and I12, while the second component is explained by I3 and I7. We can say that first component represents the profitability and growth element, while the second principal component is a Debts and Equity element.

**Rotated Component Matrix[a]**

|  | Component 1 | Component 2 |
|---|---|---|
| I1 | ,938 | ,118 |
| I2 | ,891 | ,247 |
| I3 | ,513 | ,802 |
| I4 | ,838 | ,257 |
| I6 | -,763 | ,128 |
| I12 | ,542 | -,098 |
| I7 | ,210 | -,929 |

Extraction Method: Principal Component Analysis
RotationMethod: Varimax with Kaiser Normalization

**Component Score Coefficient Matrix**

|  | Component 1 | Component 2 |
|---|---|---|
| I1 | ,268 | -,027 |
| I2 | ,240 | ,060 |
| I3 | ,066 | ,455 |
| I4 | ,223 | ,072 |
| I6 | -,243 | ,165 |
| I12 | ,173 | -,122 |
| I7 | ,165 | -,616 |

Extraction Method: Principal Component Analysis.
Rotation Method: Varimax with Kaiser

After identifying the most relevant financial ratios that describe the two principal components, ( I1, I2, I3, I4, I6, I7 and I12 ) we applied a hierarchical cluster analysis by using the Nearest Neighbour Method in order to classify the Romanian companies in two distinct clusters. It resulted a classification extremely close to the initial classification of the listed firms into "healthy" and "unhealthy" companies. Only 3 companies were miss-classified as "healthy". As a conclusion we can say that when using the Profit Margin, ROA, ROE, Profit per employee, Current Ratio, Debts on Equity and Growth rate on Total Assets variables in a cluster analysis we can reach a quite good classification of the "healthy" and "unhealthy" companies.

**CHAID CLASSIFICATION TREE**

We studied the case when using all 14 financial ratios for all the 55 listed companies, out of which I1 (Profit Margin), I2 (ROA) and I13 (Turnover growth) turned out to be most significant in the financial distress prediction problem. The following classification rules resulted: if I1< 0.04 then
                    If I2 < 0.03 => "unhealthy"
                    Else if I2> 0.03 => "healthy"
    Else if I1> 0.04 then
                    If I13 < 44.17 => "healthy"
                    Else if I13> 44.17 => "unhealthy"
The decision tree is presented below:

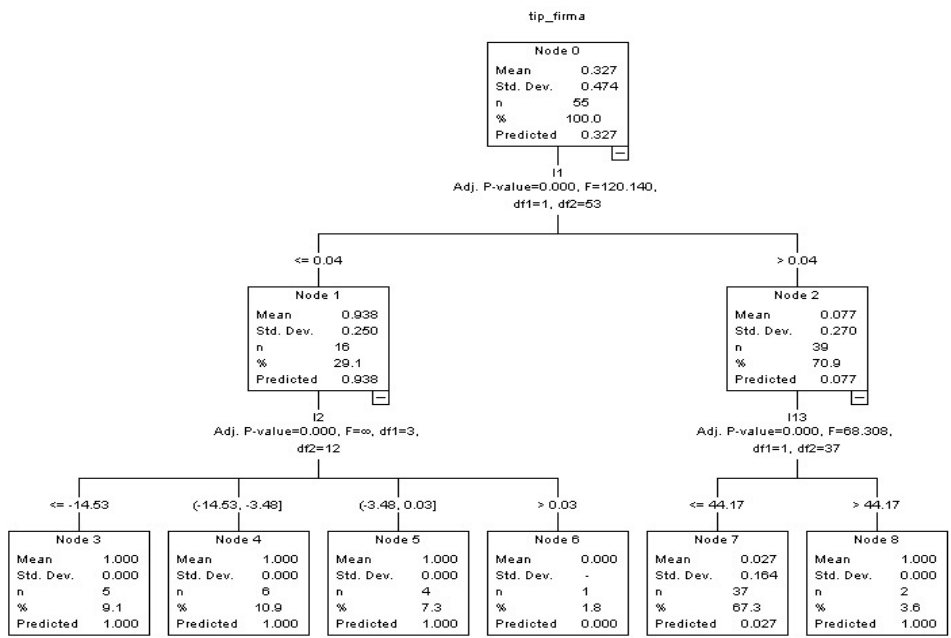

**THE LOGISTIC MODEL**

Several different models were estimated by experimentation using EViews 5.0 and having as criteria for selecting the final variable set: the high significance of the variables for the model, the correct sign of the coefficient in the model and a high level of prediction accuracy for the training sample. Out of the backward looking method, the following valid model resulted, in which the most significant financial ratios for distress prediction are I1 (Profit Margin) and I7 (Debts on Equity):

**Tip_firma= f( I1, I7)**
Dependent Variable: TIP_FIRMA
Method: ML - Binary Logit (Quadratic hill climbing)
Sample: 1 55
Included observations: 55
Convergence achieved after 10 iterations
Covariance matrix computed using second derivatives

| Variable | Coefficient | Std. Error | z-Statistic | Prob. |
|---|---|---|---|---|
| I1 | -0.828685 | 0.311621 | -2.659270 | 0.0078 |
| I7 | 0.007475 | 0.004436 | 1.685006 | 0.0920 |
| C | -1.539466 | 0.763775 | -2.015601 | 0.0438 |

| | | | |
|---|---|---|---|
| Mean dependent var | 0.327273 | S.D. dependent var | 0.473542 |
| S.E. of regression | 0.219904 | Akaike info criterion | 0.452211 |
| Sum squared resid | 2.514605 | Schwarz criterion | 0.561702 |
| Log likelihood | -9.435804 | Hannan-Quinn criter. | 0.494552 |
| Restr. log likelihood | -34.77267 | Avg. log likelihood | -0.171560 |
| LR statistic (2 df) | 50.67373 | McFadden R-squared | **0.728643** |
| Probability(LR stat) | **9.92E-12** | | |
| Obs with Dep=0 | 37 | Total obs | 55 |
| Obs with Dep=1 | 18 | | |

**5. CONCLUSIONS**

In order to identify the "healthy" and "unhealthy" Romanian listed companies for the year 2008 we applied several models and methodologies, such as the principal component analysis, a hierarchical cluster, CHAID decision tree model and the logit model. All models classified the listed companies quite good and provided relevant information of the financial ratios that better predict financial distress. The PCA and cluster analysis indicated the following variables: the Profit Margin, ROA, ROE, Profit per employee, Current Ratio, Debts on Equity and Growth rate on Total Assets, the CHAID decision tree model indicated Profit Margin, ROA and Turnover growth, while the logit model indicated Profit Margin and Debts on Equity.